\documentclass{article}
\usepackage{graphicx}
\usepackage[margin=2cm]{geometry}
\usepackage{xcolor}
\usepackage{colortbl}
\usepackage{lineno}
\title{Superconductivity and Charge-density-wave-like Transition in Th$_2$Cu$_4$As$_5$}
\author{
Qing-Chen Duan\textsuperscript{1,6}{$^{\#}$},
Shao-Hua Liu\textsuperscript{1}{$^{\#}$},
Bai-Zhuo Li\textsuperscript{1},
Jiao-Jiao Meng\textsuperscript{1},
Wu-Zhang Yang\textsuperscript{3},\\
Yi Liu\textsuperscript{2,4},
Yi-Qiang Lin\textsuperscript{2},
Si-Qi Wu\textsuperscript{2},
Jia-Yi Lu\textsuperscript{2},
Jin-Ke Bao\textsuperscript{5},
Yu-Sen Xiao\textsuperscript{1,8},
Xin-Yu Zhao\textsuperscript{2},\\
Yu-Xue Mei\textsuperscript{1},
Yu-Ping Sun\textsuperscript{1},
Dan Yu\textsuperscript{1},
Shu-Gang Tan\textsuperscript{1},
Qiang Jing\textsuperscript{1},
Rui-Dan Zhong\textsuperscript{7},\\
Yong-Liang Chen\textsuperscript{8},
Yong Zhao\textsuperscript{9},
Zhi Ren\textsuperscript{3},
Cao Wang\textsuperscript{1}\thanks{wangcao@sdut.edu.cn},
and Guang-Han Cao\textsuperscript{2,7}\thanks{ghcao@zju.edu.cn}
}
\date{}
\begin{document}

\maketitle

\begin{abstract}
We report the synthesis, crystal structure, and physical properties of a novel ternary compound, Th$_2$Cu$_4$As$_5$. The material crystallizes in a tetragonal structure with lattice parameters $a=4.0716(1)$ {\AA} and $c=24.8131(4)$ {\AA}. Its structure can be described as an alternating stacking of fluorite-type Th$_2$As$_2$ layers with antifluorite-type double-layered Cu$_4$As$_3$ slabs. The measurement of electrical resistivity, magnetic susceptibility and specific heat reveals that Th$_2$Cu$_4$As$_5$ undergoes bulk superconducting transition at 4.2 K. Moreover, all these physical quantities exhibit anomalies at 48 K, where the Hall coefficient change the sign. These findings suggest a charge-density-wave-like (CDW) transition, making Th$_2$Cu$_4$As$_5$ a rare example for studying the interplay between CDW and superconductivity.
\end{abstract}

\section{Introduction}

As widely recognized, superconductivity research encompasses two critical avenues: (1) The quest for new structural units that may harbor superconductivity preferably with higher transition temperatures, and (2) The study of the competition/cooperation between superconductivity and charge/spin-ordered states, providing insights into the superconducting mechanism. The iron-based superconductors based on the Fe$_2$As$_2$ layer serve as an excellent example, where superconductivity and antiferromagnetism engage in a competitive relationship.{\color{red}\cite{Hosono-2008-JACS, Ce1111-neutron}}. From a crystallographic perspective, the conducting Fe$_2$As$_2$ layer belong to an antifluorite-like $M_2As_2$ family, where $M$ represents transition metal elements. Over the past 15 years, researchers have unveiled the remarkable flexibility of the $M_2$As$_2$ layer, as it readily accommodates most of 3$d$ transition metal elements at the $M$ site{\color{red}\cite{Just-1996-JAC,Poettgen-2008-Z. Naturforsch. B}}. Correspondingly, the physical properties of these compounds are highly contingent on the 3$d$ elements present. For instance, compounds endowed with Cr$_2$As$_2$ or Mn$_2$As$_2$ layers exhibit metallic or semiconducting behavior with antiferromagnetism{\color{red}\cite{Park-2013-Inorg. Chem,Filsinger-2017-PRB,An-2009-PRB,Islam-2020-PRB}}. In contrast, compounds featuring Co$_2$As$_2$ layers are itinerant ferromagnet{\color{red}\cite{Yanagi-2008-PRB}}, while those housing Ni$_2$As$_2$ layers exhibit superconductivity along with Pauli paramagnetism{\color{red}\cite{Watanabe-2007-Inorg. Chem,Li-2008-PRB}}. As for the copper-based materials, say BaCu$_2$As$_2$, prior theoretical studies proposed an $sp$ metallic character, a claim subsequently supported by angle-resolved photoemission spectroscopy{\color{red}\cite{BaCu2As2-1, BaCu2As2-Singh, BaCu2As2-ARPES}}. To date, there is no solid evidence of superconductivity in compounds with single-layered Cu$_2$$Pn_2$ ($Pn$ =P, As) blocks{\color{red}\cite{BaCu2As2-1,Han-2010-JACS,Han F-2011-JACS}}. Nevertheless, superconductivity has been recently discovered in layered compounds $R_2$(Cu$_{1-x}$Ni$_x$)$_5$As$_3$O$_2$ ($R$ = La, Pr, Nd){\color{red}\cite{CXL-2532}}. This finding hints at the potential of multi-layered Cu-As structural motif to act as superconducting layers.

 In this study, we report the synthesis, structure, and physical properties of Th$_2$Cu$_4$As$_5$, a novel copper-based ternary compound. The compound adopts a tetragonal U$_2$Cu$_4$As$_5$-type structure with space group $I4/mmm$ (No. 139){\color{red}\cite{Kaczorowski-1991-JLM}}. The structure can be derived from the antifluorite-type Cu$_2$As$_2$ layers, which condense to form a double-layer slab with the composition Cu$_4$As$_3$. These Cu$_4$As$_3$ slabs are separated by layers of fluorite-type Th$_2$As$_2$, resulting in the overall composition of Th$_2$Cu$_4$As$_5$, as shown in Figure 1(c). The compound exhibits bulk superconductivity below 4.2 K. Furthermore, anomalies in electrical resistivity, magnetic susceptibility, specific heat, and Hall coefficient are observed at around 48 K, indicative of a possible charge density wave (CDW) transition. On the basis of physical property measurements and band structure calculations, we attribute the superconducting transition to the Cu$_4$As$_3$ slab, while the CDW state appears to be linked to the square As sheet in the Th$_2$As$_2$ layers. This discovery provides a rare example of superconductivity within Cu-As-layer systems, offering a new platform for studying the interplay between the CDW state and superconductivity.

\section{Results}

\begin{figure}[htbp]
	\centering
	\includegraphics[width=0.7\linewidth]{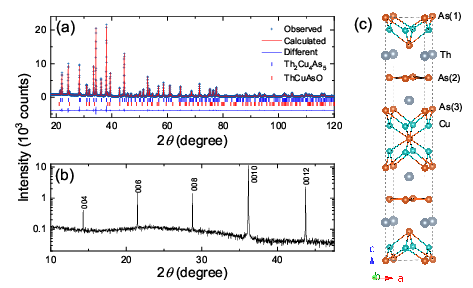}
	\caption{(a) XRD pattern and Rietveld refinement profile for polycrystalline Th$_2$Cu$_4$As$_5$ sample. (b) XRD pattern of a  single-crystalline Th$_2$Cu$_4$As$_5$ sample. (c) The unit cell of Th$_2$Cu$_4$As$_5$.}
\end{figure}

 \begin{table}[htbp]
 	\centering
 	\caption{Crystallographic Data of Th$_2$Cu$_4$As$_5$ at 300 K.}
 	\begin{tabular}{cccccc}
 \hline
 		\rowcolor{lightgray} \multicolumn{3}{c}{compounds} &  \multicolumn{3}{c}{Th$_2$Cu$_4$As$_5$} \\
 		\multicolumn{3}{c}{space group} &  \multicolumn{3}{c}{$I$4/mmm} \\
 		\multicolumn{3}{c}{$a$ ({\AA})} & \multicolumn{3}{c}{4.0716(1)} \\
 		\multicolumn{3}{c}{$c$ ({\AA})} &  \multicolumn{3}{c}{24.8131(4)} \\
 		\multicolumn{3}{c}{$V$ ({\AA}$^3$)} &  \multicolumn{3}{c}{ 411.34(3)} \\
 		\multicolumn{3}{c}{$R_{\rm wp}$ (\%)} &  \multicolumn{3}{c}{4.85\%} \\
 		\multicolumn{3}{c}{$\chi^2$ } &  \multicolumn{3}{c}{1.497} \\
 		\rowcolor{lightgray} atom & mult. & $x$ & $y$ & $z$ & $U_{\rm iso}$  \\
 		Th & 4 & 0 & 0 & 0.1567 & 0.0060 \\
 		Cu & 8  & 0 & 0.5 & 0.0527 & 0.0178 \\
 		As(1) & 2 & 0 & 0 & 0 & 0.0143 \\
 		As(2) & 4 & 0 & 0.5 & 0.25 & 0.0083 \\
 		As(3) & 4 & 0 & 0 & 0.3832 & 0.0107 \\
 		\rowcolor{lightgray}	\multicolumn{2}{c}{bond}&\multicolumn{2}{c}{multiplicity} &\multicolumn{2}{c}{bond length ({\AA})}\\
               \multicolumn{2}{c}{Th-As(2)}&\multicolumn{2}{c}{4}&\multicolumn{2}{c}{3.084} \\
 		\multicolumn{2}{c}{Th-As(3)}&\multicolumn{2}{c}{4}&\multicolumn{2}{c}{3.043} \\
 		\multicolumn{2}{c}{Cu-As(1)} &\multicolumn{2}{c}{2} &\multicolumn{2}{c}{2.422}\\
 		\multicolumn{2}{c}{Cu-As(3)} &\multicolumn{2}{c}{2}& \multicolumn{2}{c}{2.583}\\
 \multicolumn{2}{c}{As(2)-As(2)}&\multicolumn{2}{c}{4}&\multicolumn{2}{c}{2.879} \\
 		\hline
 	\end{tabular}
 \end{table}

\subsection{Crystal structure}
Figure 1(a) presents the XRD pattern and Rietveld refinement profile for the polycrystalline Th$_2$Cu$_4$As$_5$ sample, while Figure 1(b) shows the XRD pattern for a single-crystalline sample. The structural details are summarized in Table 1. The determined lattice parameters are $a=4.0716(1)$ {\AA} and $c=24.8131(4)$ {\AA}.  Compared with U$_2$Cu$_4$As$_5$, both the $a$- and $c$-axes of Th$_2$Cu$_4$As$_5$ are notably longer, which is consistent with the difference in ionic radii between Th$^{4+}$ and U$^{4+}$. Upon closer scrutiny, it becomes evident that the lattice expansion resulting from the replacement of Th$^{4+}$ with U$^{4+}$ is nearly isotropic, as indicated by the axial ratios ($c/a=6.090$ for U$_2$Cu$_4$As$_5$ and 6.094 for Th$_2$Cu$_4$As$_5$). However, the thickness of the Cu$_4$As$_3$ slab in Th$_2$Cu$_4$As$_5$ (5.797 Å) is approximately 2.9\% shorter than that in U$_2$Cu$_4$As$_5$ (5.971 Å). Thus the substitution of U$^{4+}$ with Th$^{4+}$ introduces compressive strain along the $c$-axis in the Cu$_4$As$_3$ layers. Another noteworthy structural detail in Th$_2$Cu$_4$As$_5$ is the atomic distance. It is important to consider that the Th-As(3) distance of 3.043 Å is significantly smaller than the sum of the ionic radii of Th$^{4+}$ ions (1.04 Å) and As$^{3-}$ anions (2.22 Å), suggesting orbital hybridization between Th and As(3). Simultaneously, the As-As distance within the As(2) sheet measures 2.879 Å, indicating the presence of a covalent bond. It is worth noting that similar As or Sb planes also exist in 112-type $Ln$Cu$Pn_2$ compounds ($Ln$ represents lanthanide elements, $Pn$=As, Sb) and play a decisive role in their electrical transport properties{\color{red}\cite{RCuAs2-1, RCuSb2-band}}.

\subsection{Electrical resistivity}

\begin{figure}[htbp]
	\centering
	\includegraphics[width=0.7\linewidth]{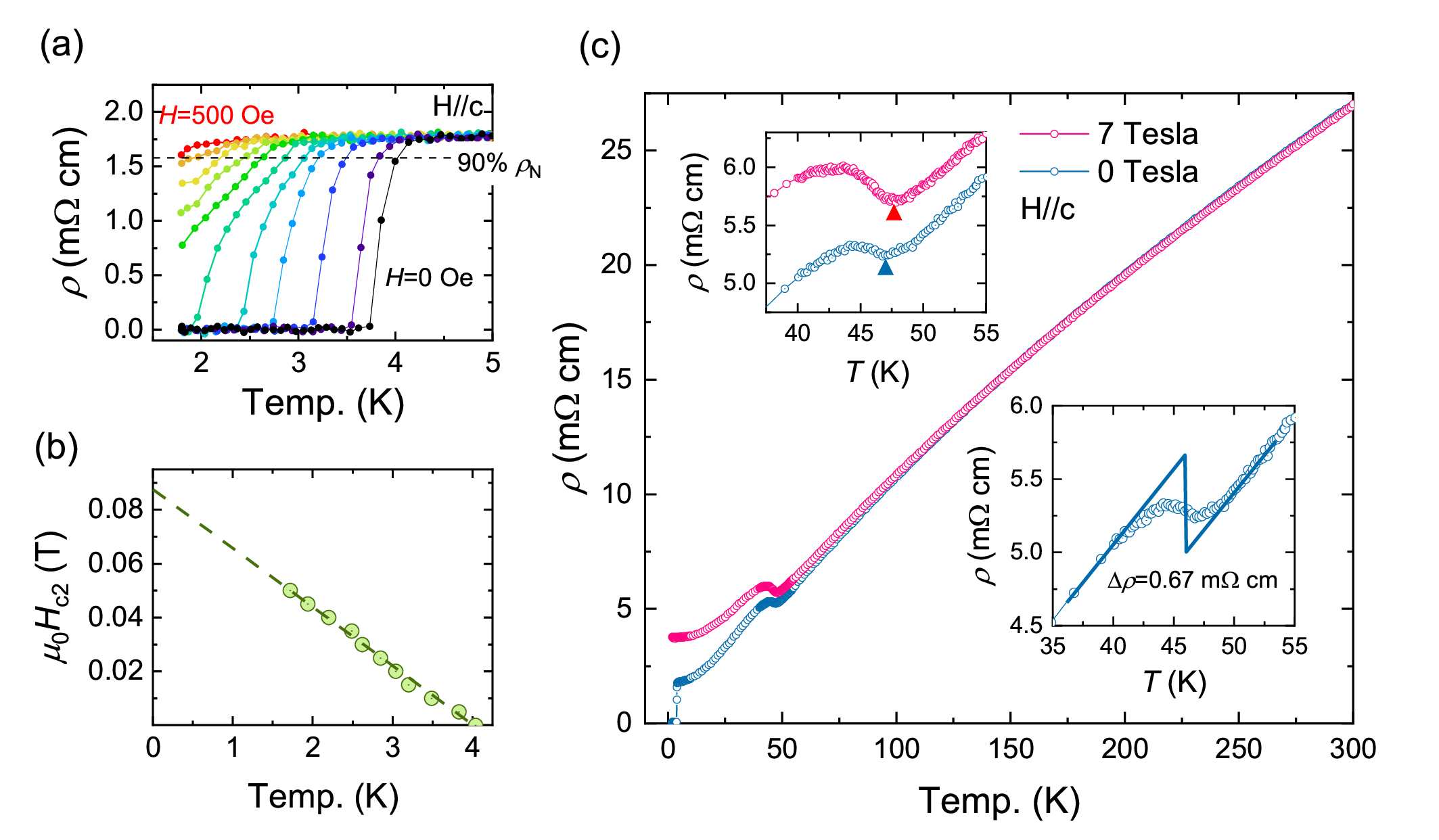}
	\caption{(a) In-plane electrical resistivity measured on a single-crystalline Th$_2$Cu$_4$As$_5$. The measurement was performed in selected magnetic fields parallel to the $c$-axis. (b) Temperature dependence of the upper critical field. (c) Normal state in-plane resistivity measured under static magnetic fields of 0 tesla and 7 tesla. The upper inset presents an enlarged view near the anomalies for both of the curves, while the lower inset depicts the magnitude of the jump in zero-field resistivity at the phase transition temperature.}
\end{figure}

Figure 2 depicts in-plane electrical resistivity data from a single-crystalline sample. The magnetic field was parallel to the $c$-axis. At zero field, the resistivity sharply decreases to zero at 4.2 K, indicating the superconducting phase transition. We define the superconducting transition temperature $T_{\rm c}$($H$) as the temperature at which the resistivity, under different magnetic fields, drops to 90\% of its normal-state value. As depicted in Figure 2(b), the temperature-dependent upper critical field exhibits linear growth upon cooling. Consequently, the upper critical field at 0 K can be linearly extrapolated to be $H_{\rm c2}$ (0) = 872 Oe, and the in-plane coherence length can be calculated as $\xi_{ab}$ = 614 {\AA} according to the relation $\mu_0$$H_{\rm c2}$(0) = $\Phi_0$/($2\pi\xi_{ab}^2$), where $\Phi_0$ is the magnetic flux quantum.

Figure 2(c) presents the $\rho$-$T$ curves spanning the entire temperature range. The zero-field resistivity measures 26.9 $\mu\Omega\cdot\rm{cm}$ at 300 K, and exhibits metallic behavior upon cooling. As the temperature approaches approximately 48 K, the resistivity undergoes a conspicuous dip, followed by a noticeable bump. This behavior typically suggests opening of an energy gap on a portion of the Fermi surface. As the temperature descends below 43 K, the $\rho$-$T$ curves reverts to metallic behavior. In the upper inset of Figure 2(c), we observe the dip shifts 0.9 K towards higher temperatures under a magnetic field of 7 tesla. Since all the data were collected using the same heating program, this shift is attributed to intrinsic physical properties. On the basis of this observation, we posit that the most likely cause of this resistivity anomaly is a CDW phase transition. Then, based on the magnitude of the resistivity jump, as illustrated in the lower inset of Figure 2(c), we estimate an 11.6\% reduction in the density of states near the Fermi surface ($N(E_F)$) due to the phase transition, assuming all the carriers have the same mobility.

\subsection{Magnetic susceptibility}

\begin{figure}[htbp]
	\centering
	\includegraphics[width=0.7\linewidth]{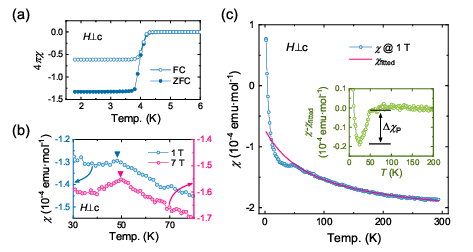}
	\caption{(a) Temperature dependent magnetic susceptibility of single-crystalline Th$_2$Cu$_4$As$_5$ sample measured in a static field of 10 Oe. The magnetic field is perpendicular to the $c$-axis. (b) The magnified view of the susceptibility near the CDW-like transition, measured at static fields of 10$^4$ Oe and $7\times10^4$ Oe, respectively.  (c) The $\chi$-$T$ curve over the entire temperature range measured at 10$^4$ Oe. The pink solid line fits the susceptibility in the range of 60 K - 300 K using the extended Curie-Weiss law $\chi=\chi_0+C/(T-\theta)$. The inset shows the deviation between the measured susceptibility and the fitting curve.}
\end{figure}

Figure 3 depicts the results of magnetic susceptibility measurements for the single-crystalline sample. The measurements were performed with the magnetic field perpendicular to the $c$-axis. As shown in Figure 3(a), the magnetic susceptibility exhibits a pronounced diamagnetic signal below 4.2 K, confirming the bulk superconductivity. Figure 3(b) presents a comparative analysis of the $\chi$-$T$ curves around 50 K under magnetic fields of 1 tesla and 7 tesla. Cusp-like anomalies are clearly visible at around 48 K, coinciding with the features observed on the $\rho$-$T$ curves. Notably, the temperature at which the cusp appears on the 7 T curve is slightly higher than on the 1 T curve (1.2 K). The increase is essentially in line with the findings from the resistivity measurements, suggesting a slight elevation in the ordering temperature for the CDW-like transition ($T_{\rm CDW}$) with the growing magnetic field. Similar behaviors, as observed in one-dimensional NbSe$_3$ and two-dimensional organic $\alpha$-(BEDT-TTF)$_2$KHg(SCN)$_4${\color{red}\cite{1D-NbSe3,2D-organic}},  have been attributed to the dominance of the orbital effect over the Pauli effect under imperfect nesting conditions{\color{red}\cite{CDW-review}}.

To gain deeper insights into the CDW-like transitions, we applied the Curie-Weiss law, $\chi=\chi_0+C/(T-\theta)$, to fit the $\chi$-$T$ curve gathered at 1 T above 60 K. The derived parameters are as follow: $\chi_0=-2.12\times10^{-4}$ $\rm{emu\cdot mol^{-1}}$, $C= 0.0079$ emu$\cdot$K$\cdot$mol$^{-1}$, $\theta=-33$ K. The negative value of $\chi_0$ indicates that the Langevin diamagnetic contribution ($\chi_{\rm L}$) surpasses the sum of the Pauli susceptibility ($\chi_{\rm P}$) and the orbital Van Vleck contribution ($\chi_{\rm VV}$). In general, both $\chi_{\rm L}$ and $\chi_{\rm VV}$ are insensitivity to a CDW transition.  Consequently, the decrease in $\chi (T)$ at 48 K can be attributed to the reduction in $\chi_{\rm P}$, indicating a decrease in $N(E_F)$. Moreover, the Curie constant corresponds to effective magnetic moments of 0.12 $\mu_{\rm B}$/Cu, providing further evidence for the negligible Cu local moments. Therefore, the Curie-Weiss behavior above 60 K can be attributed to the presence of paramagnetic impurities. We utilize the fitted $\chi$-$T$ curve as a background for subtraction. As shown in the inset of Figure 3(c), the $\Delta\chi_{\rm P}$ value associated with the transition can be estimated as 1.7$\times10^{-5}$ $\rm{emu\cdot mol^{-1}}$. We will address this issue later.

\subsection{Specific heat}

\begin{figure}[htbp]
	\centering
	\includegraphics[width=0.7\linewidth]{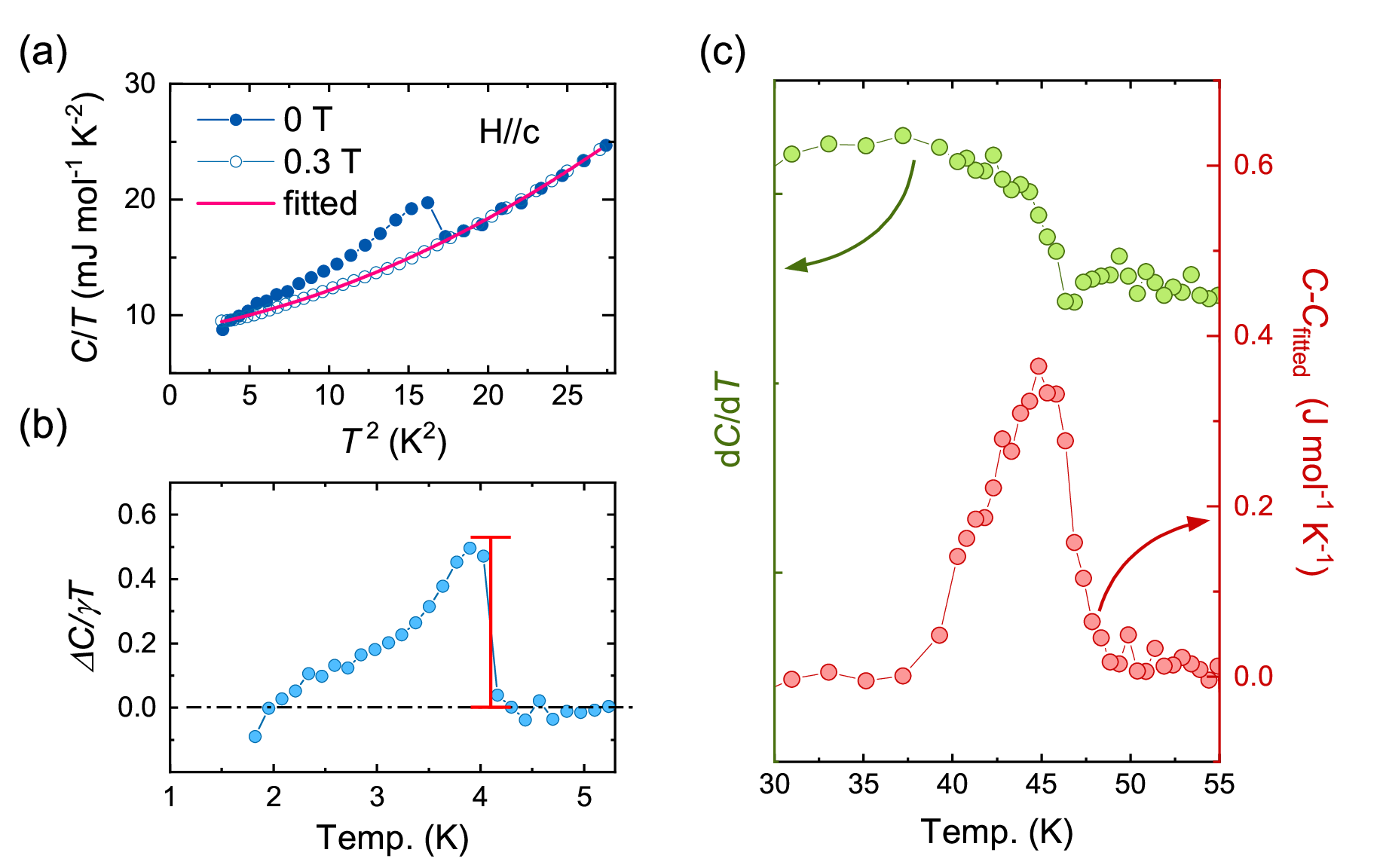}
	\caption{(a) Temperature dependent specific heat measured at static fields of 0 T and 0.3 T, respectively. (b)  The specific heat jump resulting from the superconducting transition, where $\Delta C$ is calculated as $\Delta C=C_{0 T}-C_{0.3 T}$. (c) Left axis: derivative of specific heat with respect to temperature. Right axis: specific heat values after subtracting a background, derived by polynomial fitting of the raw data within the temperature ranges 25 K $<T<$ 35 K and 55 K $<T<$ 60 K.}
\end{figure}

Figure 4 presents temperature-dependent specific heat measurements on a single crystal sample of Th$_2$Cu$_4$As$_5$. In Figure 4(a), it is evident that the zero-field specific heat exhibits a significant jump below 4.1 K, indicating a bulk superconducting transition. However, when exposed to a magnetic field of 0.3 T, the superconducting signal disappears. We fit the specific heat data below 6 K (\emph{B}=0.3 T) using the formula $C=\gamma T + \beta T^3 + \alpha T^5$, where the first term accounts for the contribution from electrons while the subsequent terms represent contributions from the lattice. The obtained parameters are $\gamma = 8.6$ mJ$\cdot$mol$^{-1}\cdot$K$^{-2}$, $\beta = 0.228$  mJ$\cdot$mol$^{-1}\cdot$K$^{-4}$, and $\alpha= 0.0132$ mJ$\cdot$mol$^{-1}\cdot$K$^{-6}$. Thus the $N(E_F)$ value can be deduced from the formula $N(E_F)=3\gamma/(\pi^2k_B^2N_A)$ (where $k_B$ is Boltzmann’s constant), and $N(E_F)$ is found to be 3.65 states$\cdot$eV$^{-1}$$\cdot$f.u.$^{-1}$. In Figure 4(b), after subtracting the specific heat under the field, a clear specific heat jump of $\Delta C/(\gamma T)= 0.53$ is observed, which is significantly smaller than the expected value of 1.43 for a weak-coupling BCS superconductor. Furthermore, it's worth mentioning that the $\Delta C/(\gamma T)$-$T$ curve below the superconducting $T_c$ cannot be fitted using an isotropic single-gap superconducting model. The origin of this phenomenon may involve multi-gap superconductivity.

Figure 4(c) shows the results of specific heat measurements in the vicinity of the CDW-like transition. By conducting a temperature derivative analysis, a distinct anomaly in d$C/$d$T$ becomes evident, which corresponds to the phase transition observed in both magnetic susceptibility and resistivity data. We performed a polynomial fitting of the specific heat raw data in the periphery of the CDW-like phase transition, and used it as a background for subtraction. This reveals a distinct specific heat peak extending from 48 K down to 37 K. Notably, the amplitude of the peak closely resembles that observed in LaAgSb$_2$, despite the significantly higher $T_{\rm CDW}$ values in LaAgSb$_2$ (207 K and 186 K) compared to Th$_2$Cu$_4$As$_5${\color{red}\cite{LaAgSb2-CDW-1,LaAgSb2-CDW-2,LaAgSb2-specific heat}}. This phenomenon challenges the linear relationship between $\Delta C$ and $T_{\rm{CDW}}$ predicted by the Ehrenfest mean-field theory{\color{red}\cite{1975-Testardi, CDW-HC jump}}. Nevertheless, the jump in specific heat for Th$_2$Cu$_4$As$_5$ implies that the CDW-like transition is intrinsic. In addition, we can relate the Sommerfeld coefficient ($\gamma$) to $\chi_{\rm P}$ using the equation $\chi_{\rm P} = 3\gamma\mu_{\rm B}^2/(\pi k_{\rm B})^2$ by assuming the Wilson ratio reaches unity. Consequently, we can estimate $\chi_{\rm P}$  well below $T_{\rm CDW}$ of Th$_2$Cu$_4$As$_5$ as $\chi_{\rm P} = 1.18 \times 10^{-4}$ $\rm{emu\cdot mol^{-1}}$. Since $\chi_{\rm P}$ can be defined as $\chi_{\rm P} = \mu_{\rm B}^2N_{\rm A}N(E_F)$, the specific heat and the drop in magnetic susceptibility ($\Delta\chi_{\rm{P}}$) enable us to estimate that the CDW-like transition results in a gap of approximately 12.7\% in the density of states near the Fermi energy. Notably, this estimate is in good agrement with the resistivity data.

\subsection{Hall effect}

\begin{figure}[htbp]
	\centering
	\includegraphics[width=0.7\linewidth]{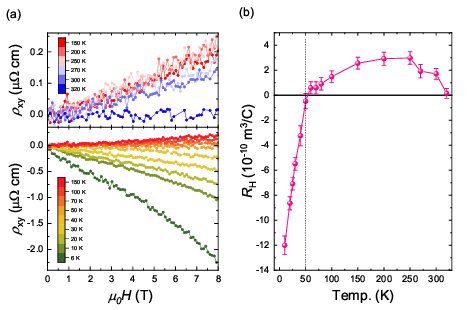}
	\caption{(a) Field-dependent Hall resistivity $\rho_{xy}$ at different temperatures. In this measurement, a polycrystalline sample was used. The $\rho_{xy}$ value was calculated as $\rho_{xy} = [\rho(+H) - \rho(-H)]/2$ to eliminate the effects of electrode misalignment. (b) Temperature dependence of the Hall coefficient $R_H$.}
\end{figure}

Figure 5(a) presents the field-dependent Hall resistivity ($\rho_{xy}$) at various temperatures. The measurement was performed with a polycrystalline sample.  The negative values of $\rho_{xy}$ obtained at temperatures below 50 K indicate that electron-type charge carriers dominate this temperature range, as evidenced by the negative Hall coefficient $R_{\rm H}$ = $\rho_{xy}/H$ displayed in Figure 5(b). Conversely, for temperatures exceeding 50 K, $R_{\rm H}$ changes from negative to positive and attains its maximum value at 250 K. After that, the $R_{\rm H}$ value decreases again with increasing temperature, ultimately approaching zero at 320 K. This behavior strongly implies a multi-band characteristic in Th$_2$Cu$_4$As$_5$, sharply distincts from copper-based LiCu$_2$P$_2$ and BaCu$_2$As$_2${\color{red}\cite{BaCu2As2-Singh,Han F-2011-JACS}}.  Additionally, the Hall coefficient exhibits a kink-like anomaly near 50 K, which also coincides with the CDW-like transition.  This observation further suggests that the CDW-like transition near 48 K may originate from Fermi surface nesting mechanisms in a multi-band system.

\subsection{Electronic structure}

\begin{figure}[htbp]
	\centering
	\includegraphics[width=0.9\linewidth]{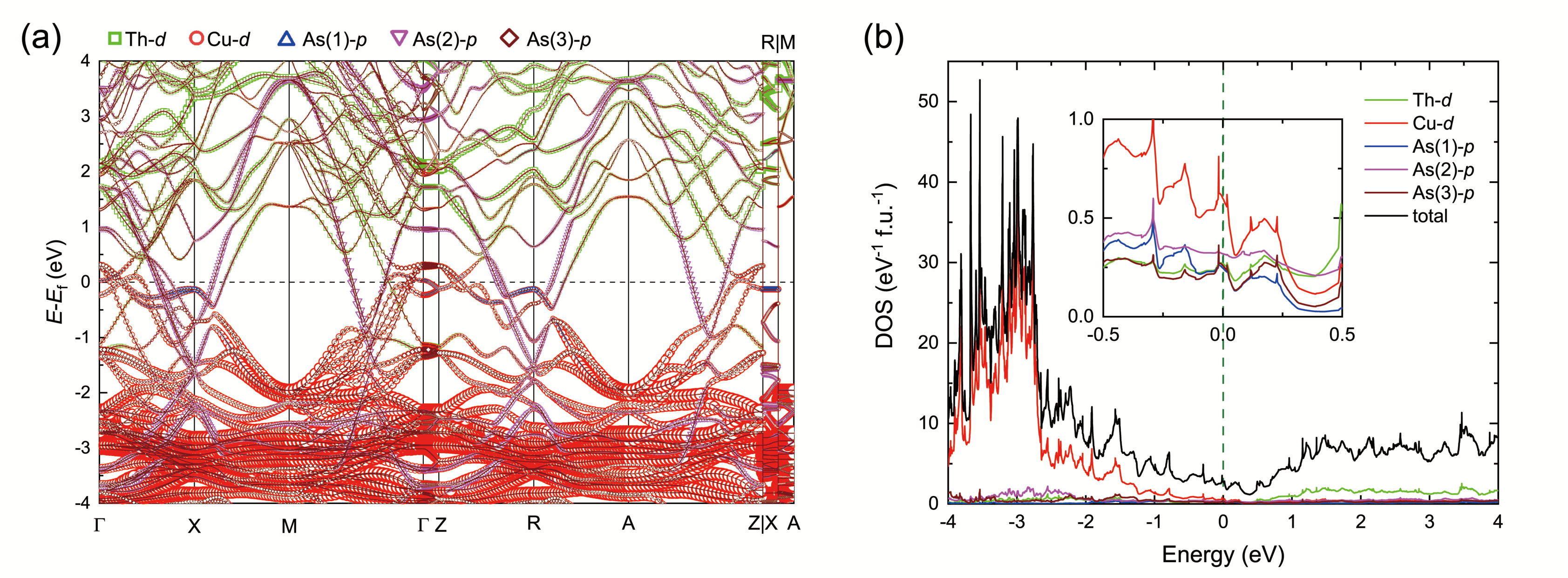}
	\caption{(a) The calculated electronic band structure. (b)The densities of states (DOS) for Th$_2$Cu$_4$As$_5$. The inset presents an enlarged view of the DOS near the Fermi energy. }
\end{figure}
The calculated band structure and DOS for Th$_2$Cu$_4$As$_5$ are presented in Figures 6(a) and 6(b). There are eight bands that intersect the Fermi level, which is consistent with the metallic nature of the material. The DOS at the Fermi energy is determined as $N(E_F)=2.76$ states$\cdot$eV$^{-1}$$\cdot$f.u.$^{-1}$. Notably, the primary contribution to the $N(E_F)$ comes from the Cu 3$d$ orbitals, in stark contrast to the 122-type BaCu$_2$As$_2$, where the 4$p$ electrons dominate the conducting band{\color{red}\cite{BaCu2As2-Singh,BaCu2As2-ARPES}}. The inset in Figure 6(b) offers an enlarged view of the DOS near the Fermi energy, showing significant hybridization between the 3$d$ orbital Cu and the 4$p$ orbitals of As(1) and As(3). Moreover, the 6$d$ orbitals of Th make a significant contribution to $N(E_F)$, consistent with the Th-As(2) bond length being shorter than expected. Thus the characteristic of Th$_2$Cu$_4$As$_5$ is reminiscent of the 112-type YCuAs$_2$, where the conducting band is contributed by both the As square sheet and the Y-Cu-As structural motif{\color{red}\cite{YCuAs2-CXH}}. In addition, given that the calculations did not consider the impact of the CDW-like transition, the calculated value of $N(E_F)$ can be revised to $N(E_F)_{\rm calc}=2.43$ states$\cdot$eV$^{-1}$$\cdot$f.u.$^{-1}$, based on the analysis of resistivity and magnetization. Thus, the discrepancy between $N(E_F)_{\rm calc}$ and the value inferred from specific heat measurements, $N(E_F)_{\rm obs}=3.65$ states$\cdot$eV$^{-1}\cdot$f.u.$^{-1}$, enables us to determine the electron-phonon coupling constant as $\lambda=0.50$, following the equation: $N(E_F)_{\rm obs}=(1+\lambda)N(E_F)_{\rm calc}$. This finding further confirms that Th$_2$Cu$_4$As$_5$ should be categorized as a weakly coupled system.

\section{Discussion}

Now let us discuss the origin of superconductivity and the CDW state in Th$_2$Cu$_4$As$_5$. Generally, both CDW and superconducting transitions lead to the opening of energy gaps on the Fermi surface. As a result, these two states are typically in competition. However, given the multi-band nature of Th$_2$Cu$_4$As$_5$, it seems unlikely that this competition occurs uniformly across all the Fermi surfaces. A more plausible scenario is that the CDW-like and the superconducting transitions might exert their influence on different bands somewhat independently. This hypothesis finds support in heat capacity data where the specific heat jump due to superconductivity is notably weaker than what BCS theory predicts. This difference can be attributed to the overestimation of $\gamma$ in $\Delta C/(\gamma T)$, as $\gamma$ is associated with the total density of states near the Fermi level.

In fact, similar coexistence behavior has been observed in Ba$_2$Ti$_2$Fe$_2$As$_4$O, where superconductivity relates to the Fe$_2$As$_2$ layers, while the CDW transition is associated with the Ti$_2$O square sheets{\color{red}\cite{Ba22241}}. As to Th$_2$Cu$_4$As$_5$, the As(2) atoms form tetragonal square sheets. Intriguingly, analogous As/Sb planes are prevalent in 112-type $LnTPn_2$ compounds, where $Ln$ represents lanthanides, $T$ denotes Cu, Ag, Au, and $Pn$ stands for As or Sb{\color{red}\cite{RCuAs2-1, LaAuSb2, RAgSb2-1,RAgSb2-2}}. Some of these compounds, such as LaAgSb$_2$ and LaAuSb$_2$, exhibit CDW states with even higher transition temperatures than Th$_2$Cu$_4$As$_5${\color{red}\cite{LaAgSb2-CDW-1, LaAgSb2-CDW-2,LaAuSb2}}. Notably, the primary contribution to $N(E_F)$ in the 112-type pnictides stems from the $p$ orbital{\color{red}\cite{112-DOS}}. On the other hand, for LaCuSb$_2$ and LaAgSb$_2$, which exhibit superconductivity with lower $T_{\rm c}$, the contribution of $d$-orbital to the conduction band is also less pronounced{\color{red}\cite{112-DOS, LaCuSb2-SC,LaAgSb2-SC}}. In short, the above-mentioned research collectively converges on an empirical rule: the transition temperatures of CDW and superconductivity are positively correlated with the contribution of $p$-orbital and $d$-orbital to $N(E_F)$, respectively. Therefore, it is reasonable to conjecture that in Th$_2$Cu$_4$As$_5$, the superconductivity may be associated with the Cu$_4$As$_3$ slabs, while the CDW phase transition could be ascribed to the As(2) square sheets.

In summary, we present the synthesis, structure, and physical properties of a novel ternary compound, Th$_2$Cu$_4$As$_5$. The structure of the compound features an antifluorite-type Cu$_4$As$_3$ slab and a tetragonal sheet of As atoms. Notably, the Cu$_4$As$_3$ layer makes the most significant contribution to the conducting band. Physical characterizations reveal a CDW-like phase transition occurring at 48 K, with superconductivity emerging below 4.2 K. Both experimental measurements and band structure calculations suggest that the CDW-like phase transition predominantly takes place within the tetragonal As(2) planes, while the superconductivity is associated with the Cu$_4$As$_3$ slab. This work marks the discovery of the first superconductor based on a double-layered anti-fluorite-type conducting layer and provides a rare instance of superconductivity coexisting with CDW state. Further research is highly needed to explore novel structures based on conducting layers of Cu$_4$As$_3$-type and to investigate the interplay between CDW and superconductivity.

\section{Methods}
The sample preparation was carried out under inert conditions. A polycrystalline sample of Th$_2$Cu$_4$As$_5$ was synthesized using Th, Cu (Alfa 99.9\%), and As (Alfa 99.99\%) as starting materials. The thorium metal was prepared as previously described{\color{red}\cite{Wang C-2016-JACS}}. The stoichiometric mixture of materials was homogenized in an agate mortar, cold-pressed into pellets, sealed in evacuated quartz ampoules and annealed at 900 °C for 50 hours. This reaction was repeated twice with intermediate regrinding. Th$_2$Cu$_4$As$_5$ single crystals were grown by the self-flux method. Th, Cu and As powders were accurately weighed in a ratio of Th:Cu:As=1:5:5 and thoroughly mixed. The mixture was loaded into a small alumina crucible, sealed in evacuated quartz ampoules. The single crystals were grown by heating the quartz ampoules to 1100 \textdegree C at a rate of 60 \textdegree C/h, held for 50 hours, and then slowly cooled to 800 \textdegree C at a rate of 2.5 \textdegree C/h. The crystals were separated from the flux by centrifugation at that temperature.

X-ray powder diffraction (XRD) was conducted at room temperature using a PANalytical X-ray diffractometer (Model EM-PYREAN) with monochromatic Cu K$\alpha$1 radiation. Structural data were obtained by Rietveld refinement using step-scan XRD data with $20^\circ\leq\theta\leq120^\circ${\color{red}\cite{McCusker-1999-JAC}}. During the refinement, we fix the atomic occupancy at 1.0 to avoid unreasonable occupancies (slightly exceeding 1.0). Magnetic measurements were performed using a Quantum Design Magnetic Property Measurement System (MPMS-XL5). The measurements of electrical resistivity and Hall resistivity measurements were carried out in a Quantum Design Physical Property Measurement System (PPMS-9 Dynacool) down to 1.8 K. Specific heat measurements were conducted on a Quantum Design PPMS-9 Evercool II using a relaxation method.

First-principles calculations were performed within the generalized gradient approximation (GGA) using the Vienna Ab initio Simulation Package (VASP). The experimental crystal structure was employed for all calculations. The plane-wave basis energy cutoff was set to 650 eV. A $\Gamma$-centered K mesh was found to be sufficient for convergence. The Coulomb and exchange parameters, U and J, were introduced using GGA + U calculations, where the parameters U and J are not independent, and the difference ($U_{\rm eff}=U-J$) is meaningful. $U_{\rm eff}$ was set to 0 eV for Cu 3$d$ and 11.0 eV for Th 5$f$ orbitals.

\section{Data Availability}
Data are available from the corresponding author upon reasonable request.

\section{Acknowledgements}
This work was supported by NSF of China (Grant Nos. 12050003, 12104260), the National Key Research and Development Program of China (2022YFA1403202), the Natural Science Foundation of Shandong Province, China (Grant Nos.ZR202211230087), and  the Key Research and Development Program of Zhejiang Province, China (Grant No. 2021C01002).

\section{Author Contributions}
C. W., and G. C. conceived and designed the project. Q. D., S, L., and J. M. prepared the sample. B. L, Y. M., and S. T. performed electrical resistivity measurement. W. Y., X. Z., and Q. J. collected the specific heat data. Y. L., and W. Y. collected Hall coefficient data. Y. X., Y. S., and R. Z. performed magnetic susceptibility data collection and analysis. Y. L., and S. W. performed the first-principles calculations. Y. C., Y. Z., and Z. R. collected the XRD data. J. B., and J. L refined the crystal structure. C. W. wrote the manuscript with contributions and critical feedback from G. C.

\section{Competing Interests}
The authors declare no competing interests.\\
\textsuperscript{1}School of Physics and Optoelectronic Engineering, Shandong University of Technology, Zibo 255000, P. R. China \\
\textsuperscript{2}School of Physics, Interdisciplinary Center for Quantum Information, and State Key Lab of Silicon Materials, Zhejiang University, Hangzhou 310058, P. R. China\\
\textsuperscript{3}School of Science, Westlake University, Hangzhou 310064, P. R. China\\
\textsuperscript{4}Department of Applied Physics, Key Laboratory of Quantum Precision Measurement of Zhejiang Province, Zhejiang University of Technology, Hangzhou 310023, P. R. China\\
\textsuperscript{5}School of Physics and Hangzhou Key Laboratory of Quantum Matters, Hangzhou Normal University, Hangzhou 311121, P. R. China\\
\textsuperscript{6}Tsung-Dao Lee Institute, and School of Physics and Astronomy, Shanghai Jiao Tong University, Shanghai 200240, P. R. China\\
\textsuperscript{7}Collaborative Innovation Centre of Advanced Microstructures, Nanjing University, Nanjing 210093, P. R. China\\
\textsuperscript{8}School of Physical Science and Technology, Southwest Jiaotong University, Chengdu 610031, P. R. China\\
\textsuperscript{9}College of Physics and Energy, Fujian Normal University, Fuzhou 350117, P. R. China\\

\#These authors contributed equally: Qing-Chen Duan, Shao-Hua Liu.\\

\end{document}